\documentclass[prl,aps,twocolumn,amsmath,amssymb]{revtex4}
\usepackage{graphicx}
\usepackage{bm}
\begin{document}

\title{Single-qubit lasing and cooling at the Rabi frequency}

\author{Julian Hauss$^{1,2}$, Arkady Fedorov$^1$, Carsten Hutter$^
{1,3}$,
Alexander Shnirman$^{1,4}$, and Gerd Sch\"on$^1$}

\affiliation{$^1$ Institut f\"{u}r Theoretische
Festk\"{o}rperphysik and DFG-Center for Functional Nanostructures
(CFN), Universit\"{a}t Karlsruhe, D-76128 Karlsruhe, Germany}
\affiliation{$^2$ Lichttechnisches Institut, Universit\"{a}t
Karlsruhe, D-76128 Karlsruhe, Germany} \affiliation{$^3$
Department of Physics, Stockholm University, AlbaNova University
Center, SE - 106 91 Stockholm, Sweden} \affiliation{$^4$ Institut
f\"{u}r Theoretische Physik, Universit\"{a}t Innsbruck, A-6020
Innsbruck, Austria}

\begin{abstract}
For a superconducting qubit driven to perform Rabi oscillations and
coupled
to a {\it slow} electromagnetic or nano-mechanical oscillator
we describe previously unexplored quantum optics effects. When the
Rabi frequency is tuned to resonance with the oscillator the latter can
be driven far from equilibrium. Blue detuned driving leads to a
population inversion in the qubit and a bi-stability with lasing
behavior of the oscillator; for red
detuning the qubit cools the oscillator. This behavior persists at
the symmetry point where the qubit-oscillator coupling is
quadratic and decoherence effects are minimized. There the system
realizes a ``single-atom-two-photon laser".
\end{abstract}

\maketitle

\begin{figure}
\includegraphics[width=7cm]
{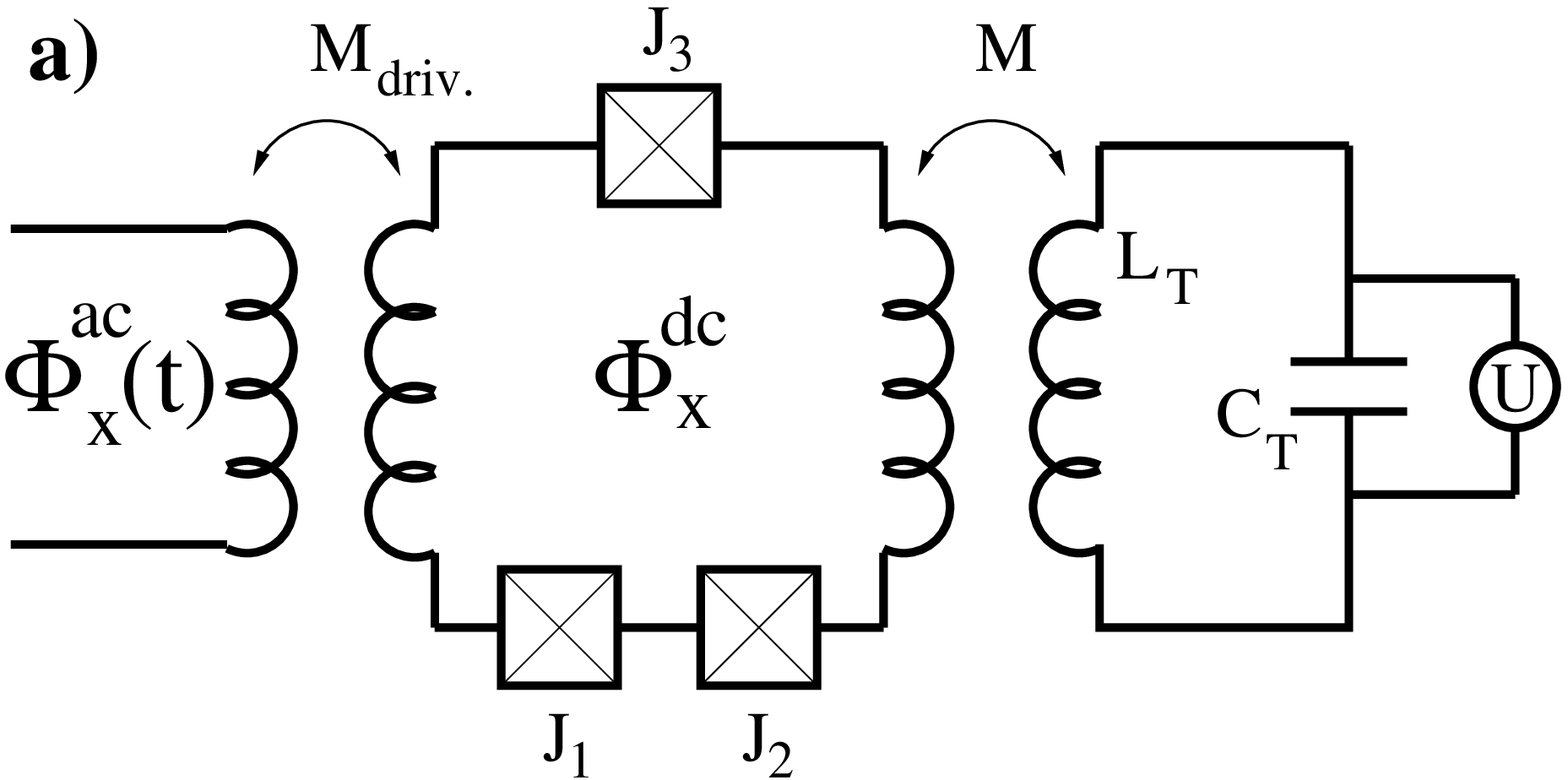}
\vskip 5mm
\includegraphics[width=4.5cm]
{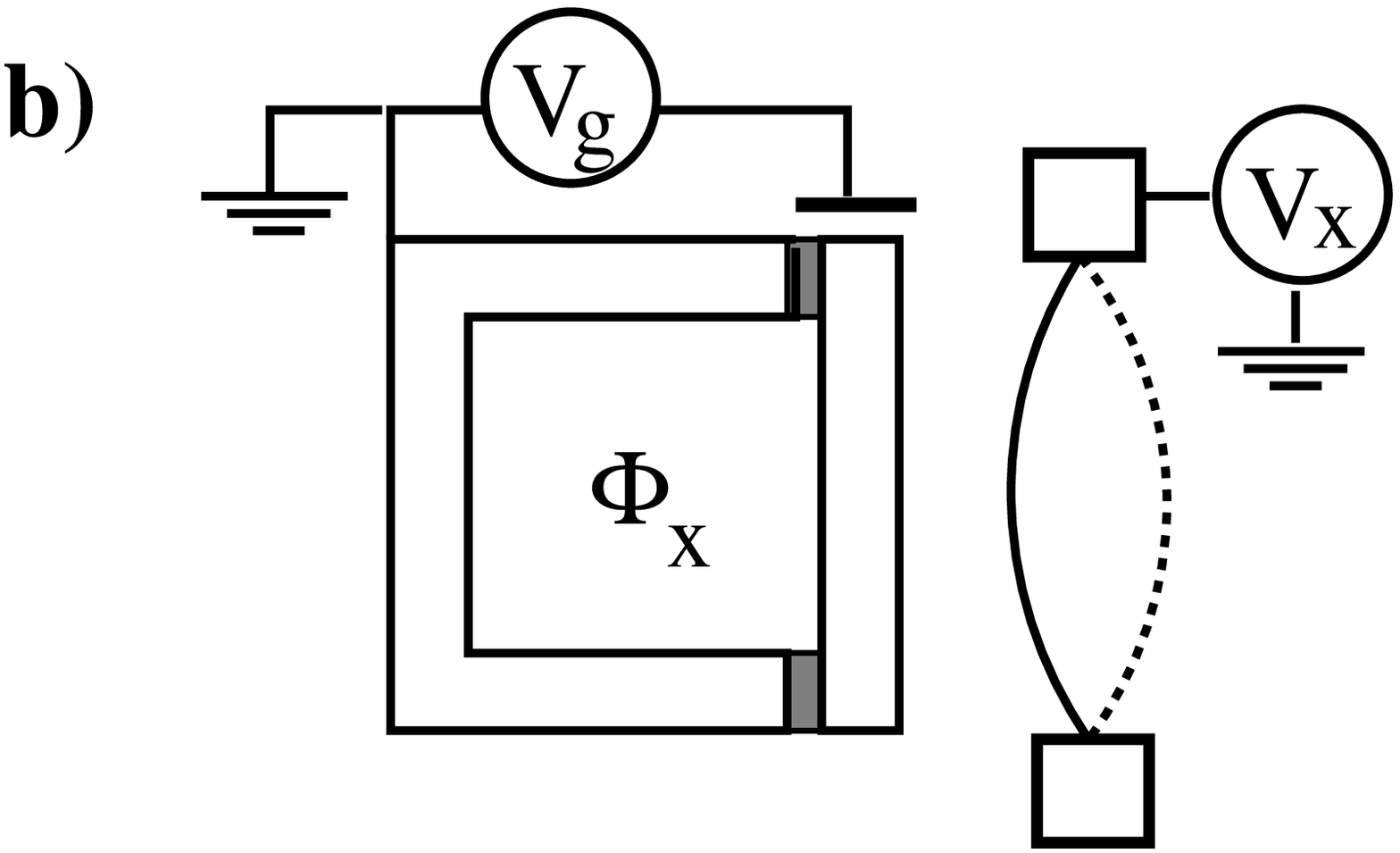} \caption{The systems. a) In the circuit QED setup of
Ref.~\cite{Jena_Rabi}
an externally driven three-junction flux qubit is coupled
inductively to an LC oscillator. b) In an equivalent setup a charge
qubit is coupled to a
mechanical resonator.} \label{fig:system}
\end{figure}

Several recent experiments on quantum state engineering with
superconducting circuits~\cite{Jena_Rabi,Wallraff_CQED,Naik} realized
concepts originally introduced in the field of quantum optics and
stimulated substantial theoretical activities ~\cite{Buisson_QED,Blais}.
Josephson qubits play the role of  two-level atoms, while
oscillators of various kinds replace the quantized light field.
Motivated by one such experiment~\cite{Jena_Rabi}, we investigate a
Josephson qubit coupled to a slow LC oscillator (Fig.~\ref{fig:system}
a) with
eigenfrequency (in the MHz range) much lower
than the qubit's energy splitting (in the GHz range),
$\omega_{T} \ll \Delta E$.
The qubit is ac-driven
to perform Rabi oscillations, and the
{\sl Rabi frequency} $\Omega_{R}$ is tuned close to resonance with
the oscillator. For this previously unexplored regime of frequencies
we study both one-photon (for $\Omega_{R} \approx \omega_{T}$)  and
two-photon (for $\Omega_{R} \approx 2 \omega_{T}$) qubit-oscillator
couplings.
The latter is dominant at the ``sweet" point of the qubit, where
due to symmetry the linear coupling to the noise sources is tuned to
zero and  dephasing effects are minimized \cite{sweet}.
When the qubit driving frequency is blue detuned,
$\delta \omega = \omega_d - \Delta E >0$, we find that the system
exhibits lasing
behavior; for red detuning the qubit cools the oscillator.
Similar behavior is expected in an accessible range of parameters for
a Josephson qubit
coupled to a nano-mechanical oscillator
(Fig.~\ref{fig:system}b), thus providing a realization of a
SASER~\cite{SASER} (Sound Amplifier by Stimulated Emission of
Radiation).

Our work is motivated by a recent experiment~\cite{Jena_Rabi} where
under the (assumed) condition $\Omega_{R} \approx \omega_{T}$ a
strong excitation of the LC oscillator was observed.
Previous attempts to explain this behavior~\cite
{Smirnov,greenberg-2005-72} did not resolve several issues:
First, according to these theories the effect should vanish at the
symmetry point, but it was observed there as well. A possible
explanation~\cite{Jena_Rabi,Smirnov}, assuming uncontrolled small
deviations from the symmetry point, is not supported by experiments.
Here we explore an alternative, namely that 2-photon processes at $
\Omega_{R} \approx 2 \omega_{T}$ which persist at the symmetry point,
are responsible for the observed effect.

The second unresolved problem is the magnitude of the effect.
The experiment showed an increase by a factor $\sim 10$ in the number
of oscillator quanta~\cite{Jena_Rabi}. The theory of Ref.~\cite
{Smirnov},
valid in the perturbative regime, predicts a much weaker effect.
Here we demonstrate that the high-frequency driving of the qubit, in
combination with the unusual resonance condition introduces
qualitatively new effects: Namely, for blue detuning of the qubit
driving a population inversion is created between the qubit dressed
states,
which are split by the Rabi frequency~\cite{PhysRev.188.1969,Zakrewski}.
Under these conditions the ac-driving provides more energy than
needed to excite the qubit, and excess energy is flowing into the
resonator.
The lasing threshold and a proper lasing state with the
characteristic line narrowing can be reached,
and the system becomes a ``single-atom-laser''~\cite{PhysRevA.46.5944}.
Similarly, for red detuning we predict strong cooling of the oscillator.

The described mechanism is strong enough to explain the observed
level of excitation of the oscillator. At this stage, using
parameters provided by Ref.~\cite{Jena_Rabi}, we have not reached
quantitative agreement with experiment, possibly because of a
combination of strong drive and low quality factor.
However, the effects should be observable in circuit QED setups of
the type described if sufficiently high $Q$-factor can be achieved.
Furthermore, our analysis applies for
low-frequency nano-mechanical resonators, where sufficiently high $Q$-
factor have already been obtained, and true lasing (sasing) should
arise.

{\it The systems } considered are shown in Fig.~\ref{fig:system}.
To be specific we first analyze the Rabi driven flux qubit coupled to an
LC-oscillator (Fig.~\ref{fig:system}a) with Hamiltonian
\begin{eqnarray}
\label{eq:Hamiltonian} H & = &
-\frac{1}{2}\,\epsilon\left(\Phi_{x}^{dc}\right){\sigma}_{z}
-\frac{1}{2}\, \Delta \, {\sigma}_{x} -
\Omega_{R0}\cos(\omega_{d}t)\, {\sigma}_{z}\nonumber\\
  &  &+\ \omega_{T}\,{a}^{\dagger}
{a} \ +\ g\, {\sigma}_{z}\left({a}+{a}^{\dagger}\right)\ .
\end{eqnarray}
The first two terms describe the qubit, with Pauli matrices
acting on the flux basis states of the qubit. The
energy bias $\epsilon(\Phi_{x}^{dc})$ between the flux states  is
controlled by a dc magnetic flux, and $\Delta$ is the
tunneling amplitude between both. The third term
accounts for the driving of the qubit by an applied ac
flux with amplitude $\Omega_{R0}$ and frequency $\omega_{d}$. The
last two terms describe the oscillator with frequency
$\omega_T=(L_T C_T)^{-1/2}$, as well as the qubit-oscillator
interaction. We estimate the coupling constant as $g\approx M I_p
I_{T,0}$, where $M$ is the mutual inductance, $I_p$ the magnitude
of the persistent current in the qubit, and
$I_{T,0}=(\omega_{T}/2L_{T})^{1/2}$ the amplitude of the
vacuum fluctuation of the current in the LC-oscillator.

After transformation to the eigenbasis of the qubit,
a Schrieffer-Wolff transformation, and a rotating wave approximation (RWA)
the Hamiltonian reads
\begin{eqnarray}
\label{eq:H_RWA} {H} & = & -\frac{1}{2}\,\Delta E\,{\sigma}_{z}
+\Omega_{R0}\cos\left(\omega_{d}t\right)
\cos\zeta\,{\sigma}_{x}
+\omega_{T}\, {a}^{\dagger}{a}\nonumber\\
&+&g\sin\zeta\,{\sigma}_{z}\left({a}+{a}^{\dagger}\right)
-\frac{g^{2}}{\Delta E}\cos^{2}\zeta\,{\sigma}_{z}
\left({a}+{a}^{\dagger}\right)^{2}
  \end{eqnarray}
where $\Delta E \equiv \sqrt{\epsilon^2 + \Delta^2}$ and $\tan\zeta=
\epsilon/\Delta$.
The derivation resembles that of Ref.~\cite{Blais} in the dispersive
regime, except that we keep terms proportional to $a^2$ and ${a^\dag}
^2$. They cannot be dropped, since the oscillator's frequency, $
\omega_T$, is much lower
than $\Delta E$. Hence, $a^2$ and ${a^\dag}^2$ evolve slowly.

Next we transform the Hamiltonian (\ref{eq:H_RWA}) into the rotating
frame,
which introduces the full Rabi frequency
$\Omega_{R}=\sqrt{\Omega_{R0}^{2}\cos^2\zeta+\delta\omega^{2}}$,
and an angle
$\tan\beta=\delta\omega/(\Omega_{R0}\cos\zeta)$ characterizing the
detuning
$\delta\omega\equiv\omega_{d}-\Delta E$. We employ a second RWA in
which we consider
$\omega_T$ and $\Omega_R \sim \omega_T$ as fast (see justification
below).
In the interaction representation with respect to the
non-interacting Hamiltonian
$H_0=(\Omega_R/2)\,{\sigma}_z+\omega_T\,
{a}^{\dagger}{a}$, we arrive at
\begin{eqnarray}
\label{HI}
H_{I}^{R}&=&g_1\left(a^{\dagger}\sigma_-e^{-i(\Omega_R-\omega_T)
t}+h.c.\right)
\label{V_I}\nonumber \\
&+&g_2\left(a^{\dagger2}\sigma_-e^{-i(\Omega_R-2\omega_T)
t}+h.c.\right)\nonumber\\
&+& g_3\left(a^{\dagger}a + a a^{\dagger}\right) \,\sigma_z\ .
\end{eqnarray}
We kept both single- and two-photon interactions with
$g_1=g\sin\zeta \cos \beta$ and $g_2=(g^2/\Delta)\cos^2\zeta \cos
\beta$, although within the RWA only one of them survives: the
first one near the resonance at $\Omega_{R} \sim \omega_{T}$,  the
second one for $\Omega_{R}\sim 2\omega_{T}$. The third term
with $g_3 = -(g^2/\Delta)\cos^2\zeta \sin \beta$
causes a frequency shift of the oscillator~\cite{greenberg-2002-66}.
All three coupling constants are much smaller than the bare $g$ in the
vicinity of the symmetry point.
In what follows we assume that the qubit is kept near the symmetry
point, i.e., $\epsilon\ll\Delta$ and $\cos\zeta\simeq1$.
Exactly at the symmetry point the single-photon coupling constant $g_1
$ vanishes, but the two-photon couplings $g_2$ and $g_3$ persist.

To account for the effects of dissipation we introduce two damping
terms in the
Liouville equation for the density operator of the system, $ \dot
\rho=-i\left[H, \rho\right]+L_Q\, \rho+L_R\,\rho$. As far
as the qubit is concerned we consider only spontaneous
emission with rate $\Gamma_0$. Hence $L_Q\, \rho=\Gamma_0\left(2
\sigma_- \rho \sigma_+ - \rho\sigma_+\sigma_-
-\sigma_+\sigma_-\rho \right)/2$. This restriction is justified as
long as the temperature is lower than the qubit's energy splitting,
and the system is biased near the degeneracy point ($\epsilon=0$)
where the additional ``pure'' dephasing is weak.
The resonator damping  can be expressed in standard form~\cite
{Gardiner}, $
L_R \, \rho=\kappa\left(N_{\rm th}+1\right)\left( 2a
\rho a^{\dagger} -a^{\dagger} a\rho -\rho a^{\dagger}a \right)/2 +
\kappa\, N_{\rm th}\left( 2 a^{\dagger} \rho a-a
a^{\dagger}\rho -\rho aa^{\dagger} \right)/2$, where
$\kappa$ denotes the resonator width and
$N_{\rm th}=1/\left[\exp(\omega_T/k_B T)-1\right]$  the thermal
average number of photons in the resonator.
Since we consider qubits
with high degree of coherence, we are motivated to study the regime
$\Gamma_0,\kappa,g_1,g_2,g_3 \ll \omega_{T},\Omega_{R}$, where the
second RWA performed above
with respect to frequencies $\omega_{T},\Omega_{R}$ is justified.

After transformations to the rotating frame and the interaction
representation,
within the RWA the qubit damping takes the form
\begin{eqnarray}\label{L_Q^R}
L_Q^R \rho_I^R&=&\frac{\Gamma_{\downarrow}}{2}\left(2 \sigma_-
\rho_I^R \sigma_+
-\rho_I^R\sigma_+\sigma_- -\sigma_+\sigma_-\rho_I^R \right)\nonumber\\
&+&\frac{\Gamma_{\uparrow}}{2}\left(2 \sigma_+ \rho_I^R\sigma_-
-\rho_I^R\sigma_-\sigma_+ - \sigma_-\sigma_+\rho_I^R \right)\nonumber\\
&+&\frac{\Gamma_{\varphi}^*}{2} \left(\sigma_z\rho_I^R\sigma_z -
\rho_I^R\right)\ .
\end{eqnarray}
Remarkably, although we started from spontaneous emission only,
the transformation to the rotating frame introduces
transition in either direction with rates $\Gamma_{\uparrow,\downarrow}=
\frac{\Gamma_0}{4}(1\pm \sin\beta)^2$ and an additional pure
dephasing term, $\Gamma_{\varphi}^*=\frac{\Gamma_0}{2}\cos^2 \beta$.
Right on resonance, where $\beta=0$, we have
$\Gamma_{\uparrow}=\Gamma_{\downarrow}$, corresponding to infinite
temperature. We further note that detuning
modifies the effective temperature in the rotating frame. For blue
detuning, $\beta > 0$, we find $\Gamma_{\uparrow} >
\Gamma_{\downarrow}$, i.e., {\it negative temperature}.

The Hamiltonian~(\ref{HI}) combined with the Liouville equation and
qubit
damping term (\ref{L_Q^R}) provides the
the theoretical description of the system.
It differs from earlier work
in the unusual resonance condition with the Rabi frequency
and the inclusion of both 1- and 2-photon processes.
The latter dominate at the symmetry point, where decoherence effects
are minimized.
Below we analyze both resonance situations
$\Omega_R\approx\omega_T$ or $\Omega_R\approx 2 \omega_T$,
and investigate the effects of blue or red
detuning, $\delta\omega = \omega_d-\Delta E$, of the
qubit driving frequency. We proceed in the frame of the standard
semi-classical approach~\cite{Reid,Gardiner} of laser physics. \\

\noindent {\sl (i) One-photon interaction}. When the Rabi
frequency is near the resonance with the oscillator, $\Omega_R\approx
\omega_T$, we can neglect in RWA in the Hamiltonian (\ref{V_I})
the term with coupling constant $g_2$.
If we - to start with - neglect fluctuations, the system is
described by Maxwell-Bloch equations for the classical variables
$\alpha=\langle a \rangle$, $\alpha^*=\langle a^{\dagger}
\rangle$, $s_{\pm}=\langle \sigma_{\pm}\rangle$, and $s_z=\langle
\sigma_z\rangle$, which can be derived from the Hamiltonian
  if all correlation functions are assumed to
factorize. The qubit variables can be adiabatically
eliminated as long as $\kappa,\, g_1\ll\Gamma_0$,
which  leads to a closed equation of motion for $\alpha$. If we
account for fluctuations, e.g., due to thermal noise in
the resonator, $\alpha$ becomes a stochastic variable obeying a
Langevin equation~\cite{Gardiner},
\begin{eqnarray}\label{dot alpha}
\dot\alpha=-\Bigg[\kappa-\frac{C}{\Gamma_\varphi + i\delta \Omega}
s_z^{st}
+  4 i g_3 s_z^{st}\Bigg]\frac{\alpha}{2}+ \, \xi(t)
  \, .
\end{eqnarray}
Here
$C \equiv 2 g_1^2$,
$s_z^{st}=-D_0/\left(1+|\alpha^2|/\tilde n_0\right)$ is the
stationary value of the population difference between the qubit
levels, and
$D_0=\left(\Gamma_\downarrow-\Gamma_\uparrow\right)/\Gamma_1$ is
the normalized difference between the rates with
$\Gamma_1=\Gamma_{\uparrow}+\Gamma_{\downarrow}$. We further
introduced the photon saturation number
$n_0=\Gamma_{\varphi}\Gamma_1/4g_1^2$ and $\tilde
n_0 \equiv n_0(1+\delta\Omega^2/\Gamma_\varphi^2)$, and the total
dephasing
rate $\Gamma_\varphi=\Gamma_1/2+\Gamma_{\varphi}^*$. The
detuning of the Rabi frequency enters in combination with a
frequency renormalization, $\delta\Omega \equiv \Omega_R-\omega_T
+ 4 g_3 |\alpha|^2$. The Langevin force due to thermal noise in the
oscillator satisfies $\langle \xi(t)\xi^*(t')\rangle = \kappa
N_{\rm th}\delta(t-t')$ and $\langle \xi(t)\xi(t')\rangle = 0$. Noise
originating from the qubit can be neglected provided the thermal
noise is strong, $\kappa N_{\rm th}\gg g_1^2/\Gamma_\varphi$.

\noindent {\sl (ii) Two-photon interaction}. The two-photon effect
dominates near the resonance condition $\Omega_R\approx
2\omega_T$. In this case in RWA we neglect in (\ref{V_I})
the term with coupling constant $g_1$.
The corresponding Langevin equation for the resonator variable
is of the same form as Eq.~(\ref{dot alpha})
but with $C \equiv 4 g_2^2|\alpha|^2$  and $s_z^{st}=-D_0/\left(1+(|
\alpha^2|/\tilde n_0)^2\right)$. The
photon saturation number is now given by $n_0=(\Gamma_{\varphi}
\Gamma_1/4g_2^2)^{1/2}$,
and $\tilde n_0 \equiv n_0(1+\delta\Omega^2/\Gamma_\varphi^2)^{1/2}$.
Again $\xi(t)$  represents  thermal noise, while noise arising from
the qubit can be neglected if $\kappa N_{\rm th}\gg g_2^2 \bar
n/\Gamma_\varphi$, where $\bar n \equiv \langle|\alpha|^2\rangle$
is the average number of photons. The relevant detuning of the Rabi
frequency for two-photon interaction is given by $\delta\Omega
\equiv \Omega_R-2\omega_T + 4 g_3 |\alpha|^2$.

\begin{figure}
\includegraphics[width=8cm]{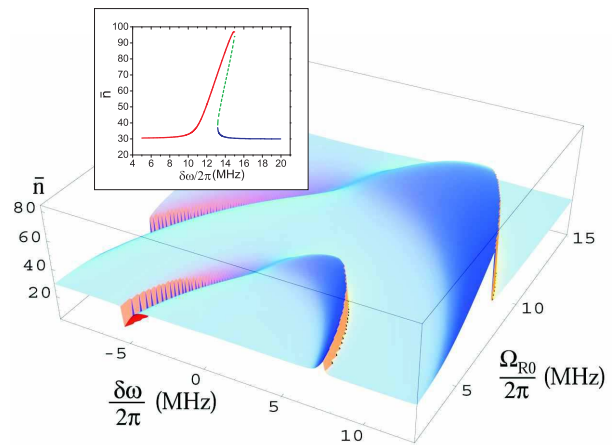}
\caption{Average number of photons
in the resonator as function of the driving detuning  $\delta\omega$
and amplitude $\Omega_{R0}$. Peaks at $\delta\omega>0$ correspond to
lasing, dips at $\delta\omega<0$ to cooling. The
inner curve is the one-photon resonance, which exists
only away from the symmetry point. (Here we assume $\epsilon =
0.01\Delta$.) The outer curve is the two-photon resonance,
which persists at $\epsilon = 0$. In domains of bi-stability the
lowest value of $\bar n$ is plotted. Sharp drops in the curves are
related to the bi-stability bifurcations. The
parameters for the qubit are $\Delta/2\pi = 1$~GHz,
$\epsilon=0.01\Delta$, and $\Gamma_0/2\pi = 125$~kHz, the frequency
and line-width of the resonator are $\omega_T/2\pi= 6$MHz and $\kappa/
2\pi = 1.7$~kHz, the coupling constant is $g/2\pi=
3.3$~MHz and the temperature of the resonator $T=10$~mK. The
inset shows the bistability of the photon number for
$\Omega_{R0}/2\pi = 7$MHz. The dashed line represents the unstable
solution.} \label{3dphoton}
\end{figure}

In Fig.~\ref{3dphoton} we summarize our main results obtained by
solving the Langevin (Fokker-Plank) equations~\cite{Reid}.
The number of photons $\bar n$ is plotted  as a function
of the detuning $\delta\omega$
of the driving frequency and driving amplitude $\Omega_{R0}$.
It exhibits sharp extrema along two curves corresponding to the
one- and two-photon resonances,
$\Omega_{R}=\omega_T-4 g_3\bar n$ and $\Omega_{R}=2\omega_T-4 g_3\bar
n$. Blue detuning, $\delta\omega>0$, induces a strong population
inversion of the qubit levels, which in resonance leads to
one-qubit lasing. In experiments the effect can be measured as a
strong increase of the photon number in the resonator
above the thermal values. On the other hand, red detuning
produces a one-qubit cooler with photon
numbers substantially below the thermal value.
Near the resonances we find regions of bi-stability illustrated in
the inset of Fig.~\ref{3dphoton}. In these regions
we expect a telegraph-like noise due to random
switching between the two solutions.

For strong population inversion, $\Gamma_\downarrow \ll
\Gamma_\uparrow$, i.e., $D_0 \sim -1$, we obtain
$\bar n \sim N_{\rm th}+ \Gamma_1/(2\kappa)$ for single-photon and
$\bar n \sim N_{\rm th}+ \Gamma_1/\kappa$ for two-photon resonances.
If $\Gamma_1/\kappa \gg N_{\rm th}$ we obtain the usual
lasing state with  Poisson distribution $p_n$ and corresponding line
narrowing.
In the opposite limit the state is almost thermal.

For comparison we plot our results using parameters as provided in
Ref.~\cite{Jena_Rabi}. As shown in Fig.~\ref{3dphoton} blue
detuning  causes a increase in the resonator
occupation number by a factor $\sim 3$, a value which is short
of the experimentally observed one. By adjusting the parameters
we can reach enhancements comparable to the observed ones. It requires
either decreasing
$\kappa$ by a factor of $\sim 4$ or increasing $\Gamma_0$ by a
factor of $\sim 4$ or a combination thereof. Further analysis and
fitting
as well as studies of higher-$Q$ circuits should help resolving the
remaining discrepancies.

So far we described a flux qubit coupled to an LC oscillator, but
our analysis applies equally to a nano-mechanical resonator
capacitively coupled to a Josephson charge qubit (see
Fig.~\ref{fig:system}b). In this case $\sigma_z$ stands for the
charge of the qubit, and both the coupling to the oscillator and
the driving are capacitive, i.e., involve $\sigma_z$. To produce
capacitive coupling between the qubit and the oscillator, the
latter is metal coated and charged by a voltage source~\cite{Naik}.
The dc component of the gate voltage $V_g$ puts the system near
the charge degeneracy point where the dephasing due to the $1/f$
charge noise is minimal. Rabi driving is induced by an ac
component of $V_g$. Realistic experimental parameters are expected
to be very similar to the ones used in the examples discussed
above, except that a much higher quality factor of the resonator
($\sim 10^5$) and a much higher number of quanta in the oscillator
can be reached. This number will easily exceed the thermal one,
thus a proper lasing state with Poisson statistics, appropriately
named SASER~\cite{SASER}, is produced. One should then observe the
usual  line narrowing with line width  given by $\kappa
N_{\rm th}/(4\bar n) \sim \kappa^2 N_{\rm th}/\Gamma_1$. Experimental
observation of this line-width narrowing would constitute a
confirmation of the lasing/sasing.

Another useful application of the considered scheme is the
cooling of the
resonator. From our analysis we conclude that a population of
order $\bar n=1$ can be reached for optimal detuning. Further
analysis is required to determine the optimum conditions.

The effect which we described here differs from the ``dressed-state
lasing'' studied earlier in quantum optics~\cite{Zakrewski}, where the
resonator is coupled to the high-frequency
Mollow~\cite{PhysRev.188.1969}
transitions with energy differences $\Delta E \pm \Omega_R$.
In the present setup the oscillator couples to
the much lower Rabi frequency (similar to Ref.~\cite{Plenio}), which can be readily tuned to
resonance with the oscillator in order to reach the lasing threshold
and a proper lasing state.

Laser-like behavior has been predicted
also for nano-mechanical resonators coupled to superconducting
single-electron transistors~\cite{blanter-2004-93}.
Our proposal differs in two major points: (i) As an active medium
we have a {\it coherently} driven qubit instead of a dissipative
transistor. While transistors usually provide
a broad band negative temperature environment for the oscillator,
frequencies of order $\omega_T$ are resolved in our situation.
Experience gained in the field of quantum optics, e.g. from the
comparison of  Doppler cooling and resolved side-band cooling, shows
that frequency resolved regimes lead to stronger effects.
(ii) We consider two-photon coupling which allows lasing at the
degeneracy
point of the qubit where the sensitivity to low-frequency noise is
substantially reduced. These facts provide sufficient flexibility in the
choice of  system parameters and should allow
reaching a proper lasing state with photon
numbers much above the thermal value and  the characteristic line-
width narrowing.

We thank E. Il'ichev, K.~C. Schwab, and M.~D. LaHaye for fruitful
discussions. The work is part of the EU IST Project EUROSQIP.

\end{document}